# GAMMA-RAY BURSTS AND BINARY NEUTRON STAR MERGERS

TSVI PIRAN
*Racah Institute for Physics*
*The Hebrew University, Jerusalem, Israel 91904*

**Abstract.** Neutron star binaries, such as the one observed in the famous binary pulsar PSR 1916+13, end their life in a catastrophic merge event (denoted here NS$^2$M). The merger releases $\approx 5 \cdot 10^{53}$ergs, mostly as neutrinos and gravitational radiation. A small fraction of this energy suffices to power $\gamma$-ray bursts (GRBs) at cosmological distances. Cosmological GRBs must pass, however, an optically thick fireball phase and the observed $\gamma$-rays emerge only at the end of this phase. Hence, it is difficult to determine the nature of the source from present observations (the agreement between the rates of GRBs and NS$^2$Ms being only an indirect evidence for this model). In the future a coinciding detection of a GRB and a gravitational radiation signal could confirm this model.

## 1. Introduction

Binary neutron star merger (NS$^2$M) is the last event in the complex evolution of some massive binary systems. A massive binary becomes a massive X-ray binary after one of the stars undergoes a core collapse and a supernova explosion. The second supernova disrupts most systems. A small fraction survives and forms a neutron star binary. There is a chance to detect a neutron star binary if one of the neutron stars is a pulsar. So far three such binaries have been detected. General Relativity predicts that the binary will emit gravitational radiation and its orbit will shrink. Taylor and Weisberg [1] measured the orbit of PSR 1916+13 and confirmed this prediction with an amazing precision. The gravitational radiation emission from the known binary pulsars is too low for a direct detection. The emission increases as the distance between the two neutron star decreases and gravitational radiation detectors such as LIGO [2] and VIRGO [3], that are being built now, will be able to detect gravitational signals from the last three minutes [4] of binary neutron stars from cosmological distances.

The outcome of the merger is most likely a rotating black hole of $\approx 2.2 - 2.8 m_\odot$ (a massive rapidly rotating neutron star, very near the upper mass limit for a rotating neutron star, cannot be ruled out yet [5]). In either case the released binding energy is $\approx 5 \cdot 10^{53}$ ergs. Shortly after the discovery of the first binary pulsar, Clark and Eardley [6] estimated that most of this energy is emitted as neutrinos, and a smaller fraction as gravitational radiation. The neutrino signal resembles a supernova neutrino burst. At present there is no way to detect such bursts from cosmological distances. Even if such a detector would have existed it would have been impossible to distinguish the rare $NS^2Ms$ neutrinos bursts from the much more frequent SN neutrino bursts.

$NS^2Ms$ conspire to release their energy in the two channels that are hardest to detect: gravitational radiation and neutrinos. However, if a small fraction of the total energy is channeled into electromagnetic energy it would be detectable. Several years ago Eichler, Livio, Piran and Schramm [7] suggested that $\nu\bar{\nu}$ annihilation ($\nu\bar{\nu} \to e^+e^-$) can convert $\approx 10^{-3}$ of the total neutrino energy to electromagnetic energy which in turn would produce a GRB (a possibility mentioned without a detailed discussion earlier by Goodman [8] Paczyński [9] and Goodman Dar and Nussinov [10]). According to this model GRBs signal the final stage in the complicated evolution of what was initially a massive binary system: the merger of a neutron star binary into a black hole (or a rapidly rotating maximal neutron star).

This idea was accepted with skepticism. At the time it was generally believed that GRBs originate from neutron star in the Galaxy (see e.g. a review by Higedron and Lingenfelter [11]). However, the BATSE detector on the COMPTON-GRO observatory have revolutionized our understanding of GRBs [12, 13]. The BATSE observations have shown that the angular distribution of GRBs is isotropic and the count distribution is incompatible with a homogeneous one. The isotropy and inhomogeneity rule out a Galactic disk population, which would be either anisotropic (if the sources are at typical distances of more than several hundred pc) or homogeneous (if the sources are at typical distances of less than several hundred pc). This leaves us with the possibility of a Galactic halo population (with a very large core radius) or a Cosmological population. While the controversy between a Galactic (mostly Galactic halo) and Cosmological populations is still going on, there is a growing consensus (see e.g. Fishman [13]) that GRBs are cosmological. With this emerging consensus the $NS^2M$ model has made the full transition from one of the least likely amongst more than a hundred GRB models to the most conservative one. It is the only model based on an independently observed phenomenon which, as we discuss shortly, takes place at a comparable rate and can easily account for the energy required.

In this lecture I discuss three issue. I examine first the GRB distribution and I show that the count distribution is compatible with a cosmological distribution (an arbitrary inhomogeneous distribution will not necessarily be compatible with a cosmological one). The analysis of the count distribution enables us to estimate the rate of GRBs, which I compare with the estimated rate of NS$^2$Ms. The agreement between the two rates clearly supports the NS$^2$M model.

I turn, next, to the issue of fireballs: The sudden release of copious $\gamma$-ray photons into a compact region, as required for a cosmological GRB, creates an opaque photon–lepton fireball due to the prolific production of electron–positron pairs. Once this was an argument against any cosmological source[14, 15]. However, Goodman [8] and Paczyński [9] have shown that the fireball expands relativistically and it releases its energy at a latter stage when it is optically thin. This optically thick phase between the energy source and the final emission stage makes the task of deciphering the GRB enigma much more difficult that what was otherwise expected.

Finally, I turn back to NS$^2$Ms. I discuss some recent numerical simulation of the hydrodynamics of the merger and I examine how does this model satisfy the constraints introduces earlier. I conclude with some open questions and predictions.

## 2. The Distribution of GRBs

### 2.1. ANGULAR DISTRIBUTION

The angular distribution of the bursts is isotropic to within the statistical errors of the sample [13]. This is clearly in agreement with a cosmological model. The isotropy sets some sever constraints on Galactic halo models [16] requiring a large homogeneous core (significantly larger than the homogeneous core expected in the dark matter distribution) and pushing the GRBs to almost inter-galactic distances. The lack of observations of GRBs from M31 sets an upper limit to the possible distance to GRBs and the combined observations begin to rule out Galactic halo models (see however [17] for a recent discussion of Galactic models).

### 2.2. COUNT DISTRIBUTION

A homogeneous population of sources in an Eucleadian space-time will have a count distribution $N(C) \propto C^{-3/2}$. Several relativistic effects influence a cosmological distribution of sources. First, space is not Eucleadian and the relation $F \propto r^{-2}$ which is the basis for $N(C) \propto C^{-3/2}$ is not valid. Second, the redshift factor, $1 + z$, changes the observed spectrum and introduces a K correction. Third, the number of counts is diluted by a $(1 + z)$ factor if the detector operates within a fixed interval $\Delta T$ that is shorter than the total duration of the burst. These effects combine to yield:

$$C(\tilde{L}, z) = \frac{\tilde{L}(1+z)^{2-\alpha}}{4\pi d_l^2(z)}, \quad (1)$$

for a detector with a fixed energy range, $\Delta E$ that operates for a fixed time interval, $\Delta t$, and sources with a count spectrum: $N(E) \propto E^{-\alpha}$. $\tilde{L}$ depends on the luminosity of the source, $L$, in the relevant energy range, $\Delta E$, on the average energy $\bar{E}$ and on the observation time, $\Delta t$: $\tilde{L} = L(\Delta E)\Delta t/\bar{E}$. $d_l(z)$ is the luminosity distance [18]. The number, $N(>C)$, of events with an observed count rate larger than $C$ is:

$$N(>C) = 4\pi \int_0^\infty n(L)dL \int_0^{z(C,L)} \frac{d_l^2}{(1+z)^3} n(z) \frac{dr_p(z)}{dz} dz \quad (2)$$

where $r_p(z)$ is the proper distance to a redshift $z$ and $z(C, L)$ is obtained by inverting Eq. 1. $n(L)$ is the luminosity function and $n(z)$ is the intrinsic rate: the number of events per unit proper volume and unit proper time at redshift, $z$. $N(C)$ depends on the cosmological parameters: $\Omega$ and $\Lambda$, and on the source parameters: $\alpha$, $n(z)$ and $n(L)$.

Cohen and Piran [19] calculated the likelihood function that the BATSE 2B (the 2B catalogue) data results from a cosmological distribution of the form given by Eq. 2 for a variety of cosmological models and source parameters. Following Kouveliotou *et. al.* [20] and Mao, Narayan and Piran [21], Cohen and Piran divided the GRB population to two sub -populations, short ($\delta t_{90} \leq 2$ s) and long ($\delta t_{90} \geq 2$ s) bursts (BATSE is more sensitive to the long bursts [21] hence it is meaningless to perform the analysis on the whole population as one group).

The maximal red shift up to which BATSE detects long bursts (for standard candles with no source evolution and spectral index $\alpha = 1.5$ [22]), is $z_{max}(long) = 2.1^{+1.1}_{-0.7}$. With estimated BATSE detection efficiency of $\approx 0.3$ this corresponds to $2.3^{-0.7}_{+1.1} \cdot 10^{-6}$ events per galaxy per year (for a galaxy density of $10^{-2}h^3$ Mpc$^{-3}$ [23]) the rate per galaxy is independent of $H_0$ and it is only weakly dependent on $\Omega$. For $\Omega = 1$ and $\Lambda = 0$ the typical energy of a burst whose observed fluence is $F_7$ (in units of $10^{-7}$ergs/cm$^2$) is $7^{+11}_{-4} \cdot 10^{50}F_{-7}$ ergs. These numbers vary slightly if the bursts have a wide luminosity function. The distance to the sources decreases and correspondingly the rate increases and the energy decreases if the spectral index is 2 and not 1.5. The rate increases and the luminosity decreases if there is a positive evolution of the rate of bursts with cosmological time.

Short bursts are detected only up to a much nearer distances: $z_{max}(short) = 0.4^{+1.1}$, again assuming standard candles and no source evolution. There is no significant lower limit on $z_{max}$ for this sub-class. The estimate of $z_{max}(short)$ corresponds to a comparable rate of $6.3^{-5.6} \cdot 10^{-6}$ events per year per galaxy and a typical energy of $3^{+39} \cdot 10^{49} F_{-7}$ ergs (note that there is no lower limit on the energy or upper limit on the rate since there is no lower limit on $z_{max}(short)$).

Luminosity functions with an effective width of up to factor of ten fit the data. This width is comparable to, and even wider than, the width of some observed luminosity distributions such as the luminosity functions of different types of supernovae.

Norris *et. al.* [24] found that the dimmest bursts are longer by a factor of $\approx 2.3$ compared to the bright ones With our canonical value of $z_{max}(long) = 2.1$ the bright bursts originate at $z_{bright}(long) \approx 0.2$. The corresponding expected ratio due to cosmological time dilation, 2.6, agrees with this measurement. If the interpretation is correct than this result confirms the cosmological model. It implies that intrinsic source evolution is insignificant and it rules out the low $z_{max}$ obtained form the combined PVO and BATSE data [25].

## 3. Fireballs

A cosmological GRB releases $10^{51}$ergs (if the energy emission is isotropic) in a very small volume. The rapid rise time observed in some bursts implies that the sources are compact with sizes, $R_i$, as small as 100km. This results in what we call a "fireball": an optically thick radiation − electron − positron plasma whose initial energy is larger than its rest mass. The initial optical depth in cosmological GRBs for $\gamma\gamma \to e^+e^-$ is:

$$\tau_{\gamma\gamma} = f_g E \sigma_T / R^2 m_e c^2 \approx 10^{19} f_g E_{i,51} R_{i,7}^{-2}, \qquad (3)$$

where $E_{i,51}$ is the initial energy of the burst in units of $10^{51}$ergs, $R_{i,7}$ is the initial radius in units of $10^7$cm and $f_\gamma$ is the fraction of primary photons with energy larger than $2m_e c^2$. Since $\tau_{\gamma\gamma} \gg 1$ the system reaches rapidly thermal equilibrium with a temperature: $T = 6.4 E_{i,51}^{1/4} R_{i,7}^{-3/4}$MeV. At this temperature there is a copious number of $e^+ - e^-$ pairs that contribute to the opacity via Compton scattering. (It is interesting that Eq. 3 yields $\tau_{\gamma\gamma} \gg 1$ even for Galactic halo objects [26]).

The huge initial optical depth prevent us from observing directly the radiation released by the source. The observed $\gamma$-rays emerge only after the fireball has expanded significantly and became optically thin. The fireball phase determines, therefore, the observational features of GRBs and it screens the specific nature of the energy source. A comparable situation is familiar in stars where the energy generated in the core leaks out through an optically thick envelope and the observed spectra is independent of the details of the energy generation mechanism at the core. Another analogous situation occurs in SNRs where the observed radiation depends just on the total energy of the ejected material and is independent of any other specific features of the source. I review here some essential physics of fireballs (see [27] for details).

### 3.1. FIREBALL EVOLUTION

Consider, first, a pure radiation fireball. Initially, the local temperature $T \gg m_e c^2$ and the opacity due to $e^+e^-$ pairs, $\tau_p \gg 1$ [8]. The fireball expands and cools until at $T_p \approx 20$ KeV, $\tau_p \approx 1$ and the photons escape freely as the fireball becomes transparent.

Astrophysical fireballs include baryonic matter in addition to radiation and $e^+e^-$ pairs. The baryons affect the fireball in two ways: The electrons associated with the baryons increase the opacity, $\tau = \tau_p + \tau_b$, delaying the escape of radiation. The baryons are also dragged by the accelerated leptons and this requires a conversion of the radiation energy into a kinetic energy. Thus, two important transitions take place as a loaded fireball evolves: The transition from optically thin to optically thick fireball, which takes place at:

$$R_\tau = (\frac{\sigma_T E}{\eta m_p c^2})^{1/2} = 6 \cdot 10^{12} \text{cm} E_{i,51}^{1/2} \eta_4^{-1/2}, \quad (4)$$

and the transition from a radiation dominated phase to a matter dominated phase which takes place at:

$$R_\eta = 2R_i \eta = 2 \times 10^{11} \text{cm } R_{i,7} \eta_4. \quad (5)$$

$\eta \equiv E_i/Mc^2$, the ratio of the initial radiation energy $E$ to the rest energy $Mc^2$ controls these transitions.

The overall outcome of the fireball depends critically on whether $R_\eta > R_\tau$ or vice versa. If $R_\eta > R_\tau$, most of the energy comes out as high energy radiation, while if $R_\tau > R_\eta$, the fireball results in a relativistic expanding shell of baryons. Energy conservation dictates that in this case $M\gamma_F \approx E$ and $\gamma_F \approx \eta$. We can classify four situations [28, 29]: (i) $\eta > \eta_{pair} = (3\sigma_T^2 E_i \sigma T_p^4/4\pi m_p^2 c^4 R_i)^{1/2} \approx 10^{10} E_{i,51}^{1/2} R_{i,7}^{-1/2}$: The effect of the baryons is negligible. The pair opacity $\tau_p$ drops to 1 and $\tau_b \ll 1$ at $T_p$ while the fireball is still radiation dominated and the radiation escapes carrying all the energy. (ii) $\eta_{pair} > \eta > \eta_b = (3\sigma_T E_i/8\pi m_p c^2 R_i^2)^{1/3} \approx 10^5 E_{i,51}^{1/3} R_{i,7}^{-2/3}$: The opacity becomes dominated by $\tau_b$. The comoving temperature decreases far below $T_p$ before $\tau$ reaches unity. The fireball is, however, still radiation dominated when $\tau_b = 1$ and the escaping radiation carries most of the energy. (iii) $\eta_b > \eta > 1$: The fireball becomes matter dominated before it becomes optically thin. The total energy is the bulk kinetic energy of extreme relativistic baryons. The final Lorentz factor is $\gamma_F \approx \eta$. (iv) $\eta < 1$: This is the Newtonian regime. The rest energy exceeds the radiation energy and the expansion is not relativistic. This is the situation, for example in supernova explosions in which the energy is deposited into a massive envelope.

A quick glance at the corresponding mass limits (the transition from case (ii) to case (iii) for $E_i \approx 10^{51}$ergs is for $M < 10^{-8} M_\odot$) reveals that case (iii) is the most likely one and even this requires a rather "clean" fireball (The transition to the Newtonian regime is at $M = 3 \cdot 10^{-3} M_\odot$). Initially such a fireball is radiation dominated and it accelerates with $\gamma \propto r$. The fireball is roughly homogeneous in its local rest frame but due to the Lorentz contraction its width in the observer frame is $\Delta r \approx R_i$, the initial size of the fireball. From $R_\eta$ onwards the baryons coasts asymptotically with $\gamma_F \approx \eta$.

## 3.2. ENERGY CONVERSION MECHANISMS

The kinetic energy of the bulk motion of the relativistic baryons can be recovered if shock waves form. The shocks could be either internal [30, 31, 32] or due to interaction with the ISM [33, 34, 35]. Variations of $\eta$ (and hence in $\gamma_F$) as a function of radius become important at $R_w$:

$$R_w \approx \eta^2 R_i \approx 10^{15} \text{cm } \eta_4^2 R_{i,7}. \qquad (6)$$

If $\gamma_F$ decreases outward than inner shells take over outer shells and internal shocks form [30]. Quite generally this is preceded by an unstable phase [32]. The shocks [30, 31] or the instability that preceded them [32] could convert a significant fraction of the kinetic energy of the baryons back to thermal energy. Since $R_w > R_\tau$ (for most reasonable parameters) this takes place in an optically thin region and the emitted photons can escape freely, producing the observed GRB.

If $\gamma_F$ increases monotonically outward than the observed pulse width increases linearly with the radius: $\Delta r \approx R_i/\gamma_F^2$ and there are no internal shock. In this case the baryons can still interact with the ISM [33, 34, 35]. Similar situation occurs in SNRs where the interaction of the ejecta with the ISM produces a shock in which the kinetic energy of the ejecta is converted into radio emission. The mean free path of a relativistic baryon in the ISM is $\approx 10^{26}$cm, hence the interaction between the baryons and the ISM cannot be collisional. However, from the existence of SNRs we can infer that a collisionless shock can form (possibly via magnetic interaction). The interaction becomes significant at $R_\gamma$ where the fireball sweeps an external mass of $M_0/\gamma_F = (E_i/\eta c^2)/\gamma_F$ and looses half of its initial momentum:

$$R_\gamma = \left[\frac{M_0}{(4\pi/3)n\gamma_F}\right]^{1/3} = 1.3 \times 10^{15}\text{cm } E_{i,51}^{1/3}\eta_4^{-2/3}n^{-1/3} \tag{7}$$

$R_w$ increases with $\eta$ while $R_\gamma$ decreases with $\eta$. Quite generally $R_\gamma < R_w$ for $\eta > 10^4$ and vice versa for $\eta < 10^4$. We should expect, therefore, that internal shocks will be important for heavily loaded "slow" fireballs while interaction with the ISM will be dominant for lightly loaded "fast" fireballs.

3.3. BEAMING AND TIMING

The fireball reaches relativistic velocities and the emitting source moves relativistically towards the observer. Thus, each observer detects radiation only from a narrow angle $\approx 1/\gamma$. This does not mean, however, that the overall GRB is beamed in a narrow angle. The angular spread of the GRB emission depends on the width of the emitting region $\Delta\theta$ which depending on the source model could be as large as $4\pi$ (but not less than $1/\gamma$).

The duration of the burst depends on several factors. The fireball appears as a narrow shell whose width could be as short at the original duration of the pulse at the source or significantly longer due to spreading of the pulse:

$$\Delta T_l \approx \begin{cases} 1.5 \times 10^{-3}\text{sec } E_{i,51}^{1/3}\eta_4^{-8/3}n^{-1/3} & \text{for } R > R_w \\ 10^{-3}\text{sec } R_{i,7} & \text{otherwise} \end{cases}. \tag{8}$$

A given observer will detect radiation from an angular scale $1/\gamma$ around his line of sight. This will lead to a typical duration of[36]:

$$\Delta T_\perp \approx R_\gamma/\gamma_F c = 5\text{sec } E_{i,51}^{1/3}\eta_4^{-5/3}n^{-1/3}. \tag{9}$$

The over all duration, $\Delta T = \max(\Delta T_l, \Delta T_\perp)$. The strong dependence of $\Delta_T$ on $\eta$ is an advantage, as it provides a possible explanation to the large variability in durations of GRBs.

4. Binary Neutron Star Mergers

Narayan, Piran and Shemi [37] and Phinney [38] estimated, using the three binary pulsars observed in the Galaxy, the rate of NS$^2$Ms as $\approx 10^{-5.5\pm.5}$/year/galaxy. This estimate is probably too high, mostly due to the fact that the current estimates of the distance to Pulsars is larger than the one used in those calculation (see e.g. Bailes [39]). Still the rate is within the range of rates quoted earlier for GRBs (recall that the rates of long and short bursts are comparable). This is a crucial, albeit indirect, evidence for the NS$^2$M model for GRBs. A small fraction of the total energy released in a NS$^2$M suffices to power a GRB at cosmological distances. Thus the NS$^2$Ms satisfies the two essential requirements from a viable model.

*Figure 1.* Logarithmic density contour lines at the end of the computation of the merger. The contours are logarithmic, at intervals of 0.25 dex

There are still many open questions. The two basic ones are: (i) What is the energy transfer mechanism into the electromagnetic channel? (ii) Is the resulting fireball sufficiently free of baryons to reach the ultra-relativistic velocities ($\gamma \gtrsim 10^2$) needed to produce a GRB?

To address these issues we [5] developed a numerical code that follows neutron star binary mergers and calculates the thermodynamic conditions of the coalesced binary. The process of coalescence, from initial contact to the formation of an axially symmetric object, takes only a few orbital periods. Some material from the two neutron stars is shed, forming a thick disk around the central, coalesced object. The mass of this disk depends on the initial neutron star spins; higher spin rates resulting in greater mass loss, and thus more massive disks. For spin rates that are most likely to be applicable to real systems, the central coalesced object has a mass of $2.4 M_\odot$, tantalizingly close to the maximum mass allowed by any neutron star equation of state for an object supported in part by rotation. Using a realistic nuclear equation of state we estimate the temperatures after the coalescence: the central object is at a temperature of $\sim 10$MeV, while shocks heat the disk to a temperature of 2-4MeV.

Fig. 1 depicts a typical density cut perpendicular to the equatorial plan. The disk is thick, almost toroidal; the material having expanded on heating through shocks. This disk surrounds a central object that is somewhat flattened due to its rapid rotation. An almost empty centrifugal funnel forms around the rotating axis and there is practically no material above the polar caps. This funnel provides a region in which a baryon free radiation-electron-position plasma could form [40]. Neutrinos and antineutrinos from the disk and from the polar caps would collide and annihilate preferentially in the funnel (the energy in the c.m. frame is larger when the colliding $\nu$ and $\bar{\nu}$ approach at obtuse angle, a condition that easily holds in the funnel). The numerical computations do not show any baryons in the funnels. The resolution of our computation is insufficient, however, to check whether the baryonic load in the funnel is low enough. The neutrinos radiation pressure on polar cap baryons can generate a baryonic wind that will load the flow. This effect depends strongly on the temperature of the polar caps [41, 42]. The estimated temperature from our computations, $\approx 2$MeV, is marginal. Our estimate is, however, least certain in low temperature regions like this.

If the core does not collapse directly to a black hole, it will emit its thermal energy as neutrinos. $\nu\bar{\nu} \to e^+e^-$ could convert $\approx 10^{-2}$ to $10^{-3}$ of the neutrino flux to electron-positron pairs and produce a GRB. The time scale for the neutrino burst is short enough to accommodate even the shortest rise times observed. Accretion of the disk onto the central object and magnetic field reconnection around the disk [31] are two additional energy sources that could power a GRB. This energy source can operate on a longer time scale independently of the dynamics of the central object.

The numerical calculations support earlier suggestions [43] that the energy release in anisotropic and that an empty funnel forms around the rotating axis of the binary system. The fireball is highly non spherical and it expands along the polar axis and forms a jet. This poses an immediate constraint on the model. If the width of the jet is $\theta$ than we observe GRBs only from a fraction $2\theta^{-2}$ of NS$^2$Ms. The rates of GRBs and NS$^2$Ms agree only if $\theta \gtrsim 0.2$ unless the rate of NS$^2$Ms is much higher than the current estimates. In fact Tutukov and Yungelson [44] suggested recently that most neutron star binaries are born with a short orbital period and their life time is too short to detect them as binary pulsars. Such systems would escape the binary pulsar statistics and could increase by one or two orders of magnitude the rate of NS$^2$M.

The duration and spectra of GRBs vary from one burst to another. The fireball phase determines both the duration and the spectra of the bursts. The source might add variability by producing fireballs with different Lorentz factors and different initial durations. Within the funnel the baryonic load varies as a function of the angular position leading to varying final Lorentz factors that, in turn, produce bursts with different durations and spectra. Another source of variability could arise from the interplay between the two energy sources in NS$^2$Ms: Neutrino annihilation and accretion energy of the disk. These mechanisms would operate on different time scale and produce different looking bursts. An additional source of diversity [5] is the distinction between systems that collapse directly to a black hole and those that undergo a longer rotating core phase. Finally, black hole-neutron star binaries are predicted to be as common as neutron star binaries[37]. Black hole neutron star mergers[45] would produce GRBs with different characteristics than NS$^2$Ms.

NS$^2$M events can take place in a variety of host systems including dwarf galaxies, or even in the intergalactic space if in the process of formation the neutron star binary is ejected from the host galaxy [31]. Hence, unlike other cosmological models it is not essential that an optical counter parts will be observed in the location of GRBs [46]. A unique prediction of the NS$^2$M model is that gravitational radiation signals from the final stages of the merger should accompany GRBs [18, 47] and vice versa (the latter is true only up to the anisotropic emission factor discussed earlier). This coincidence, which could serve to increase the sensitivity of the gravitational radiation detectors [47], would prove or disprove this model.

I thank Ehud Cohen, Tsafrir Kollat, Ramesh Narayan and Eli Waxman for many helpful discussions. This research was supported by a BRF grant to the Hebrew University and by a NASA grant NAG5-1904.

Paul C. Joss: I have a comment that is pertinent both to your talk and to that of the previous speaker. You have concentrated your attention on an essentially cosmological model for $\gamma$-ray bursts, and Gerry Fishman argued that there is no reason to invoke more than one mechanism for producing $\gamma$-ray bursts and that, in particular, Galactic neutron stars are now excluded as a source of $\gamma$-ray bursts. In fact, as shown by Gerry in his talk, $\gamma$-ray bursts encompasses a very wide range of phenomenology in terms of overall bursts duration, temporal structure, spectral shape, and mean photon energy. I'd like to suggest that it would, if anything, be somewhat surprising if this broad range of phenomenology was due to a single underlying physical mechanism and that, in particular, it may be premature to exclude Galactic neutron star as the source of a modest fraction (perhaps 10%) of observed $\gamma$-ray bursts. The current situation is reminiscent of that surrounding cosmic x-ray bursts twenty years ago. THere with a much narrower range of phenomenology, Occam's razor was frequently invoked to argue that all x-ray burst have a single physical source. This argument led theorists to speculate about a wide variety of exotic burst mechanisms, such as massive accreting black holes in the ores of Globular clusters, etc. When the work of Walter Lewin and his collaborators demonstrated that there were, in fact, two distinct types of x-ray bursts with different physical origins the theoretical situation was greatly clarified. Perhaps the same sort of clarification lies ahead in the future of our understanding of $\gamma$-ray bursts.

Reply: I completely agree that the variability of GRBs suggests that there may be two distinct populations of sources. I tried to stress in my talk that the NS$^2$M model allows for a lot of variability, but this might not be enough and one could easily hide a sub-class of about 10% of Galactic bursts amongst the observed GRBs. In fact I was led to the NS$^2$M model while considering NS$^2$Ms as gravitational radiation sources. I realized that NS$^2$M release their binding energy in practically invisible channels and I was looking for a way to convert a fraction of this energy to electromagnetic energy. I was certain that if such a mechanism exists then NS$^2$Ms would be observed via this channel. This has led me to the NS$^2$M-GRB [7, 48] model in spite of the consensus that existed at the time that GRBs originate from Galactic neutron stars. In 1990 I was asked, in a seminar on this model at the IOA, if I really believed that GRBs are not Galactic. My reply was: "GRBs are such a diverse phenomenon that I could easily imagine that a small sub-class, say of 10%, are cosmological". I am glad that now after the BATSE observations the situations has reversed and in reply to your question I would say that "GRBs are such a diverse phenomenon that I could easily imagine that a small sub-class, say of 10%, are Galactic".

W. Kundt: You mentioned the difficulty for most models to account for the delayed super-high-energy detections ($\geq$ 10GeV) from a few GRBs. In our contribution to the Untsville workshop, we explained them as the transient switch of a Geminga-like behaviour. cf. A&SS **200**, 151 (1993).

Reply: Several mechanisms, within the context of the fireball model, could produce the delayed super-high-energy detections ($\geq$ 10GeV): (i) ultra-high energy photons produced at the shock of the fireball with the ISM (the delay is between the emission from the internal shocks and the ISM shock) [49]. (ii) Production of the ultra-high energy photons from interaction of extreme relativistic protons with a dense cloud of interstellar matter in the host galaxy (the delay arises from the fact that the trajectory of the protons is not directly towards us) [50]. (iii) Production of the ultra-high energy photons from interaction of extreme relativistic protons with intergalactic protons or CMBR photons. The time delay arises from a time of flight delay of the protons.